\documentclass{svjour3}
\usepackage{amssymb}
\usepackage{amsfonts}
\usepackage{stmaryrd}
\usepackage{color}
\usepackage{epsfig}
\usepackage{graphics}
\usepackage{hyperref}

\newtheorem{assumption}{Assumption}
\newtheorem{pilot}{Pilot Example}
\newenvironment{continued}[1][continued]{\begin{trivlist}
\item[\hskip \labelsep {\bfseries #1}]}{\end{trivlist}}

\usepackage[
backend=biber,
style=apa,
sorting=nyt
]{biblatex}
\DeclareSourcemap{
  \maps[datatype=bibtex]{
    \map{
      \step[fieldset=issn, null]
      \step[fieldset=doi, null]
      \step[fieldset=url, null]
      \step[fieldset=urldate, null]
    }
  }
}

\hypersetup{
    colorlinks=true,
    urlcolor=blue,
    linkcolor=blue,
    citecolor=blue
}

\begin{document}

\title{Optimal transportation and the falsifiability of incompletely
specified economic models\thanks{This is a pre-print of an article published in \textit{Economic Theory}, volume 42, pages 355–374 (2010). The final authenticated version is available online at: \url{https://doi.org/10.1007/s00199-008-0432-y}.\\
This paper results of the combination of two manuscript circulated under the respective titles ``Part 1: Parametric restrictions on latent variables'' by the last two authors only, and ``Part 2: Moment restrictions on latent variables'' by the three authors. Financial support from NSF grant SES 0532398 is gratefully acknowledged by the three authors. Alfred Galichon' s research is partly supported by Chaire X-Dauphine-EDF-Calyon ``Finance et D\'eveloppement Durable.''}}
\author{Ivar Ekeland \and Alfred Galichon \and Marc Henry}
\institute{Ivar Ekeland
\at University of British Columbia and PIMS,
\\\email{ekeland@math.ubc.ca}
\and
Alfred Galichon
\at \'Ecole polytechnique, Paris,
\\\email{alfred.galichon@polytechnique.edu}
\and Marc Henry
\at Universit\'e de Montr\'eal, CIRANO and CIREQ,
\\\email{marc.henry@umontreal.ca}
}
\titlerunning{Falsifiability of incomplete models}
\authorrunning{Ekeland et al.}
\date{This version: September 30, 2007}

\maketitle

\begin{abstract}
A general framework is given to analyze the falsifiability of economic models based on a sample of their observable components. It is shown that, when the restrictions implied by the economic theory are insufficient to identify the unknown quantities of the structure, the duality of optimal transportation with zero-one cost function delivers interpretable and operational formulations of the hypothesis of specification correctness from which tests can be constructed to falsify the model.
\end{abstract}

\section*{Introduction}
In many contexts, the ability to identify econometric models often rests on strong prior
assumptions that are difficult to substantiate and even to analyze within the economic
decision problem. A recent approach has been to forego such prior assumptions, thus giving
up the ability to identify a single value of the parameter governing the model,
and allow instead for a set of parameter values compatible with the empirical setup. A
variety of models have been analyzed in this way, whether partial identification stems from
incompletely specified models (typically models with multiple equilibria) or from structural
data insufficiencies (typically cases of data censoring).
See \cite{Manski:2005} for a recent survey on the topic.

All these incompletely specified models share the basic fundamental structure that a set
of unobserved economic variables and a set of observed ones are linked by restrictions
that stem from the theoretical economic model. In this paper, we propose a general
framework for conducting inference in such contexts. This approach is articulated around
the formulation of a hypothesis of compatibility of the true distribution of observable
variables with the restrictions implied by the model as an optimal transportation problem.
Given a hypothesized distribution for latent variables, compatibility of the true distribution
of observed variables with the model is shown to be equivalent to the existence of a zero
cost transportation plan from the hypothesized distribution of latent variables to the true
distribution of observable variables, where the zero-one cost function is equal to one in
cases of violations of the restrictions embodied in the model.

Two distinct types of economic restrictions are considered here.
On the one hand, the case where the distribution of unobserved variables
is parameterized yields a traditional optimal transportation formulation.
On the other hand, the case where the distribution of unobserved economic variables are only
restricted by a finite set of moment equalities yields an optimization
formulation which is not a classical optimal transportation problem, but shares similar variational properties.
In both cases the inspection of the dual of the specification problem's optimization formulation
has three major benefits.

First, the optimization formulation relates the problem of falsifying
incompletely specified economic models to the growing literature on optimal transportation
(see \cite{RR:98a} and \cite{Villani:2003}),
in
particular with relation to the literature on probability metrics (see \cite{Zolotarev:97} chapter 1).
Second, the dual formulation of the optimization problem provides significant dimension reduction, thereby
allowing the construction of computable test statistics for the hypothesis of compatibility of
true observable data distribution with the economic model given. Thirdly, and perhaps most importantly,
in the case of models with discrete outcomes, the optimal transportation formulation allows
to tap into a very rich combinatorial optimization literature relative to the discrete transport
problem (see for instance \cite{PS:98}) thereby allowing inference in realistic models of industrial organization
and other areas of economics where sophisticated empirical research is being carried out.

The paper is organized as follows. The next section sets out the framework, notations and
defines the problem considered. Section~\ref{section: parametric} considers the case of parametric
restrictions on the distribution of unobserved variables, gives the optimal transportation
formulation of the compatibility of the distribution of observable variables with the economic model at hand,
and discusses strategies to falsify the model based on a sample of realizations of the
observable variables. Section~\ref{section: semiparametric} similarly
considers the case of semiparametric restrictions on the distribution of unobservable variables
and the last section concludes.

\section*{General framework and notations}
\label{section: framework and notations}
We consider, as in \cite{Jovanovic:89}, an economic model that
governs the behaviour of a collection of economic variables $(Y,U)$,
where $Y$ is a random element taking values in the Polish space
${\cal Y}$ (endowed with its Borel $\sigma$-algebra ${\cal B}_{\cal
Y}$) and $U$ is a random element taking values in the Polish space
${\cal U}$ (endowed with its Borel $\sigma$-algebra ${\cal B}_{\cal
U}$). $Y$ represents the subcollection of observable economic
variables generated by the unknown distribution $P$, and $U$
represents the subcollection of unobservable economic variables
generated by a distribution $\nu$. The economic model provides a set of restrictions on
the joint behaviour of observable and latent variables, i.e. a
subset of ${\cal Y}\times{\cal U}$, which can be represented without
loss of generality by a correspondence $G: {\cal
U}\rightrightarrows{\cal Y}$.

\begin{figure}[htbp]
\begin{center}
\vskip10pt
\includegraphics[width=12cm]{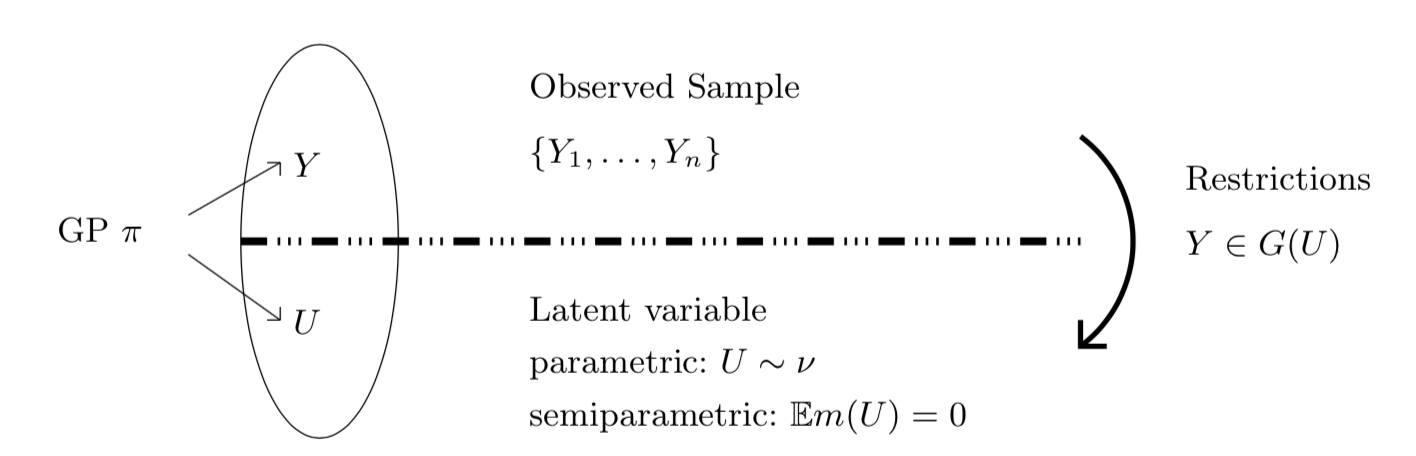} \caption{Summary of the structure.
GP stands for generating process, i.e. a joint distribution $\pi$
generating the pairs $(Y_i,U_i)$, $i=1,\ldots,n$, the first
component of which is observed.} \label{figure: structure}
\end{center}
\end{figure}

In all that follows, the
correspondence will be assumed non-empty closed-valued and measurable, i.e. $G^{-1}({\cal
O}):=\{u\in{\cal U}: G(u)\cap{\cal O}\neq\varnothing\}\in{\cal
B}_{\cal U}$ for all open subset ${\cal O}$ of ${\cal Y}$.
A measurable selection of a measurable correspondence $G$ is a measurable function $g$ such that
$g\in G$ almost surely, and Sel$(G)$ denotes the collection of measurable selections of $G$ (non-empty by
the Kuratowski--Ryll-Nardzewski selection theorem).
We shall denote by $c(y,u)$ a cost of transportation,
i.e. a real valued function on $\mathcal{Y}\times\mathcal{U}$. For any set $A$, we denote
by $1_A$ its indicator function, i.e. the function taking value $1$ on $A$ and $0$ outside of $A$.
$\mathcal{M}(\mathcal{Y})$ (resp. $\mathcal{M}(\mathcal{U})$) will denote the set of Borel probability measures on $\mathcal{Y}$ (resp. $\mathcal{U}$) and $\mathcal{M}(P,\nu)$ will denote the collection of Borel probability measures on $\mathcal{Y}\times\mathcal{U}$
with marginal distributions $P$ and $\nu$ on $\mathcal{Y}$ and $\mathcal{U}$ respectively.
We shall generally denote by $\pi$ a typical element of $\mathcal{M}(P,\nu)$.
For a Borel probability measure $\nu$ on $\mathcal{U}$ and a measurable correspondence $G: \mathcal{U}\rightrightarrows\mathcal{Y}$, we denote by $\nu G^{-1}$ the set function that to a set $A$ in $\mathcal{B}_\mathcal{Y}$ associates $\nu(G^{-1}(A))=\nu \left( \left\{ u\in \mathcal{U}:G\left( u\right) \cap
A\neq \varnothing \right\} \right)$. Note that the set function $\nu G^{-1}$ is a {\em Choquet capacity functional} (see for instance \cite{Choquet:53}). The {\em Core} of a Choquet capacity functional $\nu G^{-1}$, denoted
$\mathrm{Core}(\nu G^{-1})$ is defined as the collection of Borel probability measures set-wise dominated by $\nu G^{-1}$, i.e. $\mathrm{Core}(\nu G^{-1})=\{Q\in\mathcal{M}(\mathcal{Y}): \forall A\in\mathcal{B}_\mathcal{Y}, Q(A)\leq\nu G^{-1}(A)\}$. In the terminology of cooperative games, if $\nu G^{-1}$ defines a transferable utility game, $\nu G^{-1}(A)$ is the utility value or worth of coalition $A$ and the Core of the game $\nu G^{-1}$ is the collection of undominated allocations (see \cite{Moulin:95}).

\begin{example}[Social interactions with small groups]\label{example: social interactions}
To illustrate this framework, consider the following model of discrete choice with social interactions, in the spirit of \cite{Manski:93},
\cite{BD:2001}, but with the special feature that the interaction networks are small, so that multiple equilibria are more pervasive. Consider the variable $Y_i=1$ if individual $i$ smokes and $0$ otherwise.
Suppose the utility of individual $i$ from smoking is given by
$Y_i\left[-U_i+\sum_{j\in\mathcal{F}(i)}Y_j\right]$,
where  $\mathcal{F}(i)$ is the set of individuals that are directly connected to $i$ in the social network (a graph which is given as a primitive of the problem), $U_i$ is individual $i$'s idiosyncratic disutility of smoking. Consider for instance the following simple configuration for the social network. There are three individuals A, B and C, connected
in a line network A--B--C, so that $\mathcal{F}(A)=\mathcal{F}(C)=\{B\}$ and $\mathcal{F}(B)=\{A,C\}$.
The following are the pure strategy Nash equilibria of the game, which define the equilibrium correspondence, hence the correspondence $G$.
    \begin{itemize} \item If $U_B>2$ or ($U_B>1$ and ($U_A>1$ or $U_C>1$)) or ($U_A>1$ and $U_C>1$), then ``nobody smokes'' is the unique equilibrium. So $G((U_A,U_B,U_C))=\{(0,0,0)\}$. \item If ($U_A<1$ and $U_B<1$ and $U_C>1$) then there are two equilibria, either ``nobody smokes'' or ``A and B smoke'' (and symmetrically if the roles of A and C are reversed). So $G((U_A,U_B,U_C))=\{(0,0,0),(1,1,0)\}$.
    \item If ($U_A<1$ and $U_B<2$ and $U_C<1$) then ``everybody smokes'' and ``nobody smokes'' are both equilibria.
    So $G((U_A,U_B,U_C))=\{(0,0,0),(1,1,1)\}$. \end{itemize}
Hence, the set of observable outcomes is $\mathcal{Y}=\{(0,0,0)$, $(0,1,1)$, $(1,1,0)$, $(1,1,1)\}$. $P$ is the true distribution of equilibria in a population of identical networks (true frequencies of elements in $\mathcal{Y}$), and $\nu$ is the distribution of idiosyncratic disutilities of smoking.
\end{example}

\begin{example}[Diamond-type search model]\label{example:Diamond}
Suppose there are $N$ players searching for trading parters. Player $i$ exerts effort
$Y_i\in[0,1]$ with cost $C(Y_i)$ to find a partner. A trader's probability of finding a partner is
proportional to their own effort and the sum of other traders' efforts. Hence, the payoff function
is \[\pi_i(Y)=\epsilon Y_i\sum_{j\ne i}Y_j-C(Y_i),\] where $\epsilon$ is
the gains of trade observed by the players before making their effort choice, but not by
the econometrician, who only knows it is distributed according to distribution $\nu$, which is absolutely continuous with respect to Lebesgue measure. Assuming the cost function is increasing in effort, $x=0$ is an equilibrium, and so is
$Y=(\alpha(\epsilon),\ldots,\alpha(\epsilon))$, where $\alpha(\epsilon)$ satisfies $C'(\alpha)=\alpha(N-1)\epsilon$.
In this case, $\mathcal{Y}=[0,1]^N$ and the equilibrium correspondence is
$G(\epsilon)=\{(0,\ldots,0)^t,(\alpha(\epsilon),\ldots,\alpha(\epsilon))^t\}$. Note that since both equilibrium are perfectly correlated, this is equivalent to the simplified formulation where $\mathcal{Y}=[0,1]$ and $G(\epsilon)=
\{0,\alpha(\epsilon)\}$.\end{example}

\begin{figure}[htbp]
\begin{center}
\includegraphics[width=12cm]{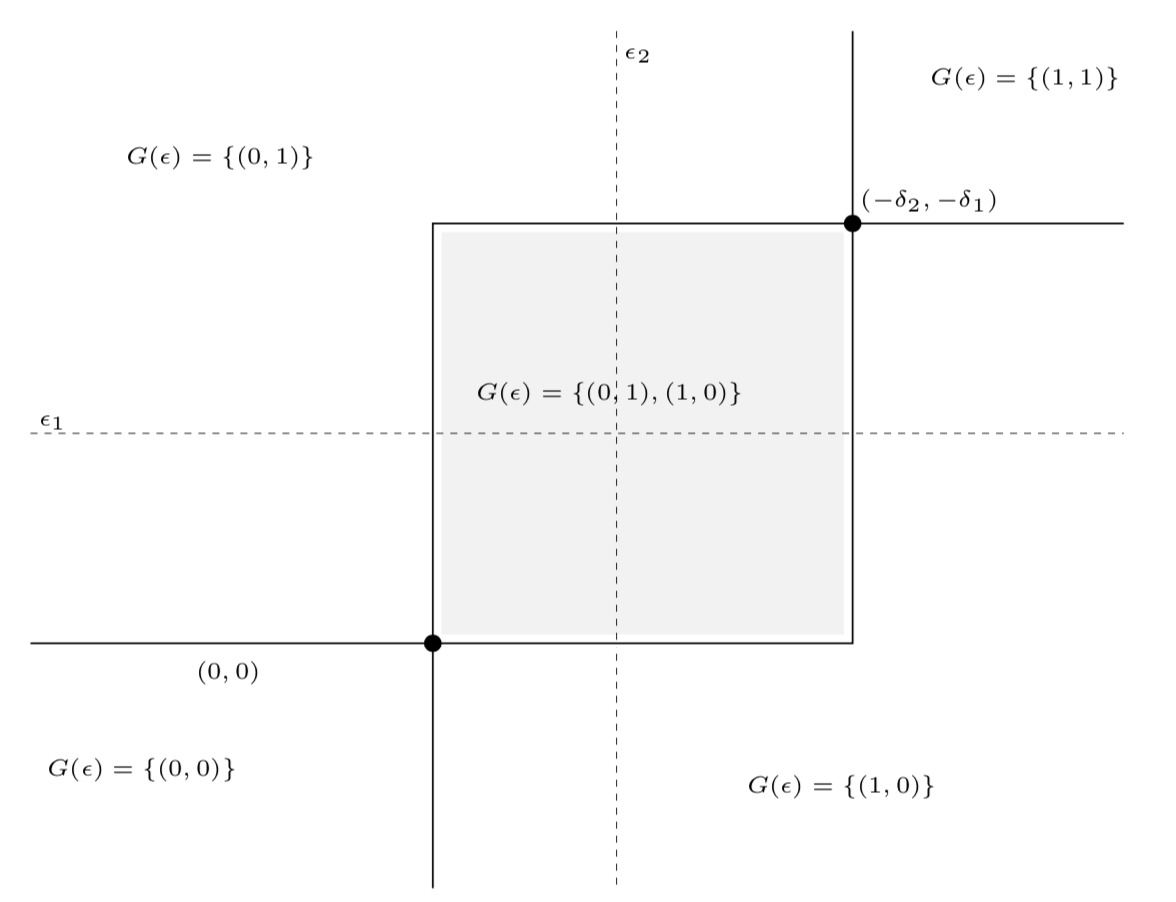}
\caption{{\scriptsize Equilibrium correspondence in example~\ref{example:Tamer}.}} \label{figure:BR22ET}
\end{center}
\end{figure}

\begin{example}[Oligopoly entry models]\label{example:Tamer}
A leading example of the framework above is that of empirical models of oligopoly entry,
proposed in \cite{BR:90} and \cite{Berry:92}, and considered
in the framework of partial identification by \cite{Tamer:2003}, \cite{ABJ:2003},
\cite{BT:2006}, \cite{CT:2006} and \cite{PPHI:2004} among others.
For illustration purposes, we describe the special case of this framework extensively studied
in \cite{Tamer:2003}, \cite{BT:2006} and \cite{CT:2006}.
Two firms are present in an industry, and a firm decides to enter the market if it makes a non negative profit in a pure strategy Nash equilibrium. $Y_{i}$ is firm $i$'s strategy, and it is equal to $1$ if firm $i$
enters the market, and zero otherwise. $Y$ denotes the vector $(Y_{1},Y_{2})$
of strategies of both firms. In standard notation, $Y_{-i}$ denotes the vector of strategies
of firm $j=3-i$. In models of oligopoly entry, the profit $\pi_{i}$
of firm $i$ is allowed to depend
on strategies $Y_{-i}$ of the other firm, as well as on a profit shifter $\epsilon_{i}$ that is observed by both firms but not by the econometrician, and a vector of unknown structural parameters $\theta$.
Profit functions are supposed to have the following linear form $\pi_{im}=\delta_{-i}Y_{-i}+
\epsilon_{i}$, where the unobserved profit shifters are distributed according to a known distribution and where parameters $\delta_{1,2}$ are given. Hence, $Y_{i}=1$ if $\delta_{-i}Y_{-i}+
\epsilon_{i}\geq0$ and zero otherwise. As noted in \cite{Tamer:2003},
if monopoly profits are larger than duopoly profits, i.e. $\delta_i<0$,
for $i=1,2$, and if $0\leq\epsilon_{i}\leq-\delta_{-i}$, $i=1,2$, then
there are multiple equilibria, since the model predicts either $Y_{1}=1$ and $Y_{2}=0$ or $Y_{1}=0$ and $Y_{2}=1$.
The set of possible outcomes
is $\mathcal{Y} = \{(0,0)$, $(0,1)$, $(1,0)$, $(1,1)\}$, and the correspondence $G$ is given in figure~\ref{figure:BR22ET}.
\end{example}

We are interested in characterizing restrictions on the distribution of
observables induced by the model, in order to devise methods to falsify the
model based on a sample of repeated observations of $Y$.
We shall successively consider two leading cases of this framework. First the case
where the distribution $\nu$ of unobservable variables is given by the economic model, and second, the case where a finite collection of moments of the distribution
$\nu$ of unobservable variables are given by the economic model.

The general principle we shall develop here in both parts is therefore the
following. We want to test the compatibility of a \textit{reduced-form model}%
, summarized by the distribution $P$ of an observed variable $Y$, with a
\textit{structural model}, summarized by a set $\mathcal{V}$ of
distributions $\nu $ for the latent variable $U$. Two leading cases will be
considered for the set $\mathcal{V}$: the parametric case, where $\mathcal{V}
$ contains one element $\mathcal{V=}\left\{ \nu \right\} $, and the
semiparametric case, where the distributions $\nu $ in $\mathcal{V}$ are
specified by a finite number of moment restrictions $\mathbb{E}_{\nu }\left[
m_{i}\left( U\right) \right] =0$.

The restriction of the model defines compatibility between outcomes of the
reduced-form and the structural models: such outcomes $u$ and $y$ are
compatible if and only if the binary relation $y\in G\left( u\right) $ holds
(this relation defines $G$).

Now we turn to the compatibility of the probabilistic models, namely of the
specification of distributions for $U$ and $Y$. The models $Y\sim P$ and $%
U\sim \nu \in \mathcal{V}$ are compatible if there is a joint distribution $%
\pi $ for the pair $\left( Y,U\right) $ with respective marginals $P$ and
some $\nu \in \mathcal{V}$ such that $Y\in G\left( U\right) $ holds $\pi $
almost surely. In other words, $P$ and $\mathcal{V}$\ are compatible if and
only if%
\[
\exists \nu \in \mathcal{V},\exists \pi \in \mathcal{M}\left( P,\nu \right)
:\Pr\nolimits_{\pi }\left\{ Y\notin G\left( U\right) \right\} =0.
\]%
In the sequel we shall examine equivalent formulations of this compatibility
principle, first in the parametric case and then in the semiparametric case.

\section{Parametric restrictions on unobservables}
\label{section: parametric}

Consider first the case where the economic model consists in the correspondence $G:\mathcal{U}\rightrightarrows
\mathcal{Y}$ and the distribution $\nu$ of unobservables. The observables are fully characterized
by their distribution $P$, which is unknown, but can be estimated from data. The question of compatibility of
the model with the data can be formalized as follows:
Consider the restrictions
imposed by the model on the joint distribution $\pi$ of the pair $(Y,U)$:
\begin{itemize} \item Its marginal with respect to $Y$ is $P$,
\item Its marginal with respect to $U$ is $\nu$, \item The
economic restrictions $Y\in G(U)$ hold $\pi$ almost
surely.
\end{itemize}
A probability distribution $\pi$ that satisfies the restrictions above may or
may not exist. If and only if it does, we say that the distribution $P$ of
observable variables is compatible with the economic model $(G,\nu)$.

\begin{definition}\label{definition: compatibility}
A distribution $P$ is compatible with the model $(G,\nu)$ for $(Y,U)$ if there exists a
probability distribution $\pi$ for the vector $(Y,U)$ with marginals $P$ and $\nu$ such that
$\pi(\{Y\in G(U)\})=1$. \end{definition}

\subsection{Optimal transportation formulation}
\label{subsection: optimal transportation}

This hypothesis of compatibility
has the following optimization interpretation. The distribution $P$ is compatible
with the model $(G,\nu)$ if and only if
$$\exists\pi\in{\cal M}(P,\nu): \int_{\mathcal{Y}\times\mathcal{U}}1_{\{y\notin G(u)\}}d\pi(y,u)=0,$$
and thus we see that it is equivalent to the existence of
a zero cost transportation plan for the problem of transporting mass
$\nu$ into mass $P$ with zero-one cost function $c(y,u)=1_{\{y\notin G(u)\}}$
associated with violations of the restrictions implied by the model.

\begin{figure}[htbp]
\begin{center}
\vskip10pt
\includegraphics[width=12cm]{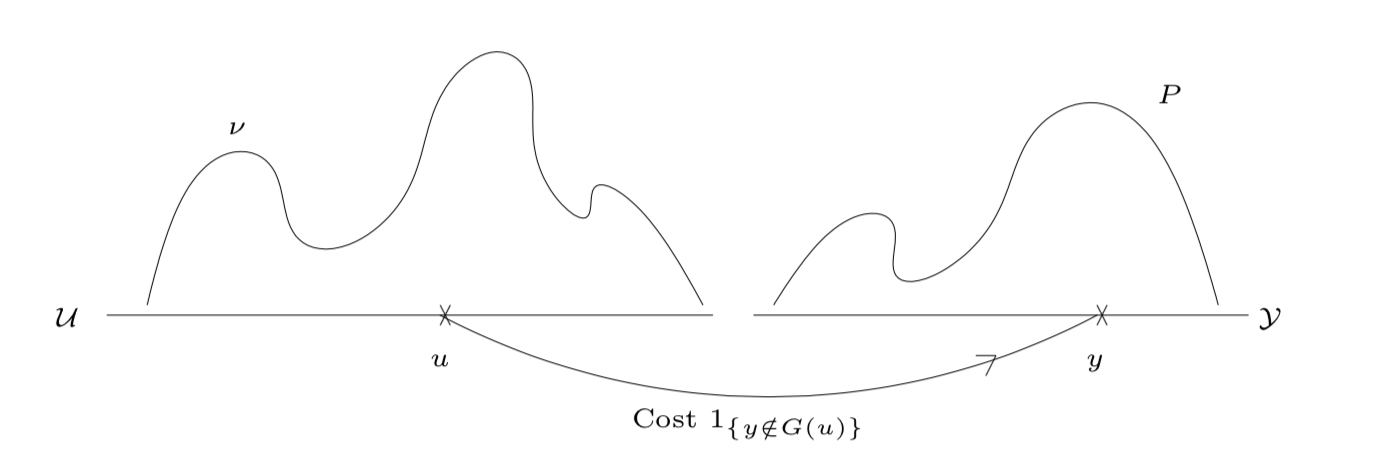} \caption{Transportation plan from mass distribution $\nu$
on $\mathcal{U}$ to mass distribution $P$ on $\mathcal{Y}$ with cost of transportation
equal to $1$ if the restrictions are violated, and $0$ otherwise.} \label{figure: transport}
\end{center}
\end{figure}

The two dual formulations of this optimal transportation problem are the following:
\begin{eqnarray*}
&(\mbox{P})& \inf_{\pi\in{\cal M}(P,\nu)} \int_{{\cal Y}\times{\cal U}} 1_{\{y\notin G(u)\}}d\pi(y,u)\\
&(\mbox{D})& \sup_{f(y)+h(u)\leq 1_{\{y\notin G(u)\}}} \int_{\cal Y} f\;dP + \int_{\cal U} h\;d\nu\end{eqnarray*}

Through applications of optimal transportation duality theory,
it can be shown that the two programs are equal and that the infimum in $(\mbox{P})$
is attained, so that the compatibility hypothesis of definition~\ref{definition: compatibility}
is equivalent to $(\mbox{D})=0$, which in turn can be shown to be equivalent to
\begin{eqnarray}\sup_{A\in{\cal B}_{\cal Y}} [P(A)-\nu(G^{-1}(A))]=0,
\label{equation: dual with sets}\end{eqnarray}
using the zero-one nature of the cost function to specialize
the test functions $f$ and $h$ to indicator functions of Borel sets.
Note that it is relatively easy to show necessity,
since the definition of compatibility implies that ${Y\in A}\Rightarrow U\in G^{-1}(A)$, so
that $1_{\{Y\in A\}}\leq1_{\{U\in G^{-1}(A)\}}$, $\pi$-almost surely.
Taking expectation, we have $\mathbb{E}_\pi(1_{\{Y\in A\}})\leq
\mathbb{E}_\pi(1_{\{U\in G^{-1}(A)\}})$, which yields $P(A)\leq\nu(G^{-1}(A))$.
The converse relies on the duality of optimal transportation (see theorem 1.27 page 44 of \cite{Villani:2003} and \cite{GH:2006d} for details). Note also that in the particular
case where the spaces of the observed and latent variables are the same $%
\mathcal{Y}=\mathcal{U}$ and $G$ is the identity function $G\left( u\right)
=\left\{ u\right\} $, then (\ref{equation: dual with sets}) defines the Total Variation metric between $P
$ and $\nu $. When $\mathcal{Y}=\mathcal{U}$ and $G\left( u\right) =\left\{
y\in \mathcal{Y}:d\left( y,u\right) \leq \varepsilon \right\} $, the above
duality boils down to a celebrated theorem due to Strassen (see section 11.6 of \cite{Dudley:2002}). A closely related result was proven by Artstein in \cite{Artstein:83}, Theorem 3.1, 
using an extension of the marriage lemma.

\begin{continued}[Example \protect\ref{example: social interactions} continued:]
In the social network example, the set of equilibria was $\mathcal{Y}=\{(0,0,0), (0,1,1), (1,1,0), (1,1,1)\}$. Let
$p_{000}$ denote the true frequency of equilibrium $(0,0,0)$, $p_{011}$ that of $(0,1,1)$, $p_{110}$ that of $(1,1,0)$
and $p_{111}$ that of $(1,1,1)$. The set of predicted combinations of equilibria is $\{\{(0,0,0)\},$ $\{(0,0,0),(1,1,0)\},$
$\{(0,0,0),(1,1,0)\},$ $\{(0,0,0),(1,1,1)\}\}$. Call $q_\omega$ the likelihood of equilibrium combination $\omega$
predicted by the model, so that for instance $q_{000,110}=\nu\{u\in\mathcal{U}: G(u)=\{(0,0,0),(1,1,0)\}\}$. The previous discussions shows that $(p_{000}, p_{011}, p_{110}, p_{111})$ is compatible with the model $(G,\nu)$ if and only if for all subset $A$ of $\mathcal{Y}$, $P(A)\leq\nu G^{-1}(A)$. Take the subset $\{(0,1,1),(1,1,0)\}$ for instance, the constraint to check is \[p_{011}+p_{110}\leq\nu G^{-1}(\{(0,1,1),(1,1,0)\})=q_{000,011}+q_{000,110},\] the $\nu$ probability of the region of $\mathcal{U}$ with incidence of the equilibrium $(0,1,1)$ or $(1,1,0)$.
\end{continued}

\begin{continued}[Example \protect\ref{example:Diamond} continued:]
In the case of the search model, it can be shown further that compatibility of the model with the distribution of observed equilibria is equivalent to $P([0,y]\leq\nu([0,\alpha^{-1}(y)])$ and $P([y,1]\leq\nu([\alpha^{-1}(y),\alpha^{-1}(1)])$ for all $y\in[0,1]$.\end{continued}

\begin{continued}[Example \protect\ref{example:Tamer} continued:]
In the case of the duopoly entry game, the model is compatible with distribution $P$ of observable variables
(indicator of presence of firms in the market) if and only if the
16 inequalities $P(A)\leq \nu(G^{-1}(A))$, or in a different notation
$\mathbb{P}(Y\in A)\leq\mathbb{P}(G(\epsilon)\cap A\ne\varnothing)$ for all sets $A$ in $2^\mathcal{Y}$.
\end{continued}

\subsection{Game theoretic interpretation of the specification problem}
The optimal transportation of the specification problem at hand leads to an interpretation of the latter
as a game between the Analyst and a malevolent Nature. This highlights connections between partial identification and robust decision making (in \cite{HS:2001}) and ambiguity (in \cite{MMR:2006}). As above, $P$ and $\nu$ are given. In the special case where we want to test whether the true functional relation between observable and unobservable variables is $\gamma_0$ (i.e. the complete specification problem), and where $P$ and $\nu$
are absolutely continuous with respect to Lebesgue measure, the optimal transportation
formulation of the specification problem involves the minimization over the set of joint probability measures
with marginals $P$ and $\nu$ of the integral $\int 1_\{y\ne\gamma_0(u)\}d\pi(y,u)$. The latter can be written as the minimax
problem \[\min_\tau\max_V\int [1_{\{\tau(u)\ne\gamma_0(u)\}}-V(\tau(u))]d\nu(u)+\int V(y)dP(y).\]

This yields the interpretation as a zero-sum game between the Analyst and Nature, where the Analyst pays Nature the amount
\begin{eqnarray}\int [1_{\{\tau(u)\ne\gamma_0(u)\}}-V(\tau(u))]d\nu(u)+\int V(y)dP(y).\label{equation:transfer}\end{eqnarray}
$P$ and $\nu$ are fixed. The Analyst is
asked to propose a plausible functional relation $y=\tau(u)$ between observed and latent variables, and Nature
chooses $V$ in order to maximize transfer (\ref{equation:transfer}) from the Analyst.
This transfer can be decomposed into two terms. The first term $\int V(y)dP(y)-\int V(\tau(u))d\nu(u)$ is a punishment for guessing the wrong distribution: this term can be arbitrarily large unless $P=\nu\tau^{-1}$. The second term, $\int 1_{\{
\tau(u)\ne\gamma_0(u)\}}d\nu(u)$ is an incentive to guess $\tau$ close to the true functional relation $\gamma_0$ between $u$ and $y$.

The value of this game for Nature is equal to $T(P)=\inf \{\mathbb{P}(\tau(U)\ne\gamma_0(U)):\;U\sim\nu,\;\tau(U)\sim P\}$ and is independent of who moves first. This follows from the Monge-Kantorovitch duality. Indeed, if Nature moves first and plays $V$, the Analyst will choose $\tau$ to minimize $\int \left(1_{\{\tau(u)\ne\gamma_0(u)\}}-V(\tau(u))\right)d\nu(u)$. Denoting
$V^\ast(u)=\inf_y\{1_{\{y\ne\gamma_0(u)\}}-V(y)\}$, the value of this game for Nature is \[
\sup_{V^\ast(u)+V(y)\leq1_{\{y\ne\gamma_0(u)\}}}\int V^\ast(u)d\nu(u)+\int V(y)dP(y).\] If, on the other hand,
the Analyst moves first and plays $\tau$, then Nature will receive an arbitrarily large transfer if $P\ne\nu\tau^{-1}$, and a transfer of $\int 1_{\{\tau(u)\ne\gamma_0(u)\}}d\nu(u)$ independent of $V$ otherwise. The value of the game for Nature is therefore $\inf \{\mathbb{P}(\tau(U)\ne\gamma_0(U)):\;U\sim\nu,\;\tau(U)\sim P\}$. The Monge-Kantorovitch duality states precisely that the value when Nature plays first is equal to the value when Analyst plays first.

Finally, we have an interpretation of the set of observable distributions $P$ that are compatible with the model
$(G,\nu)$ as the set of distributions $P$ such that the Analyst is willing to play the game, i.e. such that the value of the game is zero for some functional relationship $\gamma_0$ among the selections of $G$.

\subsection{Test of compatibility}
\label{subsection: parametric compatibility}

We now consider falsifiability of the incompletely specified model
through a test of the null hypothesis that $P$ is compatible with
$(G,\nu)$. Falsifying the model in this framework corresponds to the
finding that a sample $(Y_1,\ldots,Y_n)$ of $n$ copies
of $Y$ distributed according to the unknown true distribution $P$
was not generated as part of a sample $((Y_1,U_1),\ldots,(Y_n,U_n))$
distributed according to a fixed $\pi$ with marginal $\nu$ on ${\cal
U}$ and satisfying the restrictions $Y\in G(U)$ almost surely. Using the results of the previous section,
this can be expressed in the following equivalent ways.

\begin{proposition}
The following statements are equivalent:
\begin{eqnarray*}&\mbox{(i)}& \mbox{The observable distribution } P \mbox{ is compatible with the model } (G,\nu),\\
&\mbox{(ii)}&\inf_{\pi\in{\cal M}(P,\nu)} \int_{{\cal Y}\times{\cal U}} 1_{\{y\notin G(u)\}}d\pi(y,u)=0,\\
&\mbox{(iii)}&\sup_{A\in{\cal B}_{\cal Y}} [P(A)-\nu(G^{-1}(A))]=0.\end{eqnarray*}\end{proposition}

Call $P_n$ the empirical distribution, defined by $P_n(A) = \sum_{i=1}^{n} 1_{\{Y_i\in A\}}/n$
for all $A$ measurable, and form the empirical analogues of the conditions above as
\begin{eqnarray*}
&(\mbox{EP})& \inf_{\pi\in{\cal M}(P_n,\nu)} \int_{{\cal Y}\times{\cal U}} 1_{\{y\notin G(u)\}}d\pi(y,u)\\
&(\mbox{ED})& \sup_{A\in{\cal B}_{\cal Y}} [P_n(A)-\nu(G^{-1}(A))].\end{eqnarray*}
Note first that by the duality of optimal transportation, the empirical primal (EP)
and the empirical dual (ED) are equal. In the case $\mathcal{Y}\subseteq\mathbb{R}^{d_y}$,
\cite{GH:2006d} propose a testing procedure
based on the asymptotic treatment of the feasible statistic \[T_n=\sqrt{n}\sup_{A\in\mathcal{C}_n}
[P_n(A)-\nu G^{-1}(A)],\hskip20pt\mbox{with }\;\mathcal{C}_n=\{(-\infty,Y_i],(Y_i,\infty):\;i=1,\ldots,n\}.\]
More general families of test statistic for this problem can be derived from the following observation:
consider the total variation metric defined by
\begin{eqnarray*}d_{{\mathrm TV}}(\mu_1,\mu_2)=\sup_{A\in{\cal B}_{\cal
Y}}(\mu_1(A)-\mu_2(A))\end{eqnarray*} for any two probability
measures $\mu_1$ and $\mu_2$ on $(\mathcal{Y}, \cal{B}_\mathcal{Y})$, and
\[d_{TV}\left( P,\mathcal{Q}\right) =\inf_{Q\in \mathcal{Q}
}d_{TV}\left( P,Q\right) \] for a probability measure $P$ and a set of probability measures $\mathcal{Q}$.
\cite{GH:2008} derive conditions under which
the equalities
\begin{eqnarray*}d_{{\mathrm
TV}}(P_n,\mbox{Core}( \nu G^{-1}))&=&\inf_{ g\in{\mathrm Sel}(
G)}\sup_{A\in{\cal B}_{\cal Y}} (P_n(A)-\nu
g^{-1}(A))\\&=&\sup_{A\in{\cal B}_{\cal Y}} \inf_{ g\in{\mathrm
Sel}( G)}(P_n(A)-\nu g^{-1}(A))\\&=&\sup_{A\in{\cal B}_{\cal
Y}}(P_n(A)-\nu( G^{-1}(A)))\end{eqnarray*} hold, so that the empirical dual
is equal to the total variation distance between the empirical distribution
$P_n$ and Core$(\nu G^{-1})$. Hence, (ED) yields a family of test statistics
$d(P_n,\mbox{Core}(\nu G^{-1}))$,  for the falsification of the model $(G,\nu)$, where $d$
satisfies $d(x,A)=0$ if $x\in A$ and $1$ otherwise.

Alternatively, a family of statistics can be derived from the empirical primal (EP) if the
0-1 cost is replaced by $d$ as above, yielding the statistics \[\inf_{\pi\in{\cal M}(P_n,\nu)} \int\!\!\!\int_{{\cal Y}\times{\cal U}} d(y,G(u)) d\pi(y,u)\] generalizing goodness-of-fit statistics based on the Wasserstein distance
(see for instance \cite{BCMR:99}).

\subsection{Computational aspects of the transportation formulation}
In addition to producing families of test statistics, hence inference strategies, for partially identified structures,
the optimal transportation formulation has clear computational advantages. First of all, efficient algorithms
for the computation of the optimal transport map rely on both primal and dual formulations of
the optimization problem. More specifically, in cases with discrete observable outcomes, the Monge-Kantorovitch
optimal transportation problem reduces to its discrete counterpart, sometimes called the Hitchcock problem
(see \cite{Hitchcock:41}, \cite{Kantorovich:42} and \cite{Koopmans:49}). This problem
has a long history of applications in a vast array of fields, and hence spurred the development of many
families of algorithms and implementations since \cite{FF:57}. The optimal transportation
formulation therefore allows the development of procedures for testing incomplete structures and estimating partially identified parameters that are vastly more efficient than existing ones (see for instance \cite{GH:2008a} for
the efficient computation of the the identified set in discrete games).

\section{Semiparametric restrictions on unobservables}
\label{section: semiparametric}

As before, we consider an economic model that
governs the behaviour of a collection of economic variables $(Y,U)$.
Here, $Y$ is a random element taking values in the Polish space
${\cal Y}$ (endowed with its Borel $\sigma$-algebra ${\cal B}_{\cal
Y}$) and $U$ is a random vector taking values in
${\cal U}\subseteq\mathbb{R}^{d_u}$. $Y$ represents the subcollection of observable economic
variables generated by the unknown distribution $P$, and $U$
represents the subcollection of unobservable economic variables
generated by a distribution $\nu$. As before, the economic model provides a set of restrictions on
the joint behaviour of observable and latent variables, i.e. a
subset of ${\cal Y}\times{\cal U}\,$ represented by the measurable
correspondence $G: {\cal
U}\rightrightarrows{\cal Y}$. The distribution $\nu$ of the unobservable
variables $U$ is now assumed  to satisfy a set of moment
conditions, namely
\begin{eqnarray}\mathbb{E}_{\nu}(m_i(U)) = 0,\hskip10pt m_i:\,{\cal
U}\rightarrow\mathbb{R},\;\;i=1,\ldots,d_m \label{equation: moment
conditions}\end{eqnarray} and we denote by $\mathcal{V}$ the
set of distributions that satisfy (\ref{equation: moment
conditions}), and by $\mathcal{M}(P,\mathcal{V})$ the collection of Borel probability measures
with one marginal fixed equal to $P$ and the other marginal belonging to the set $\mathcal{V}$.
Note that a limit case of this framework, where an infinite collection of moment conditions
uniquely determines the distribution of unobservable variables, i.e. when $\mathcal{V}$ is a singleton,
we recover the parametric setup, with a classical optimal transportation formulation as in section~\ref{section: parametric}.

\begin{example}[Model defined by moment inequalities.]
\label{example: moment inequalities} A special case of the
specification above is provided by models defined by moment
inequalities.
\begin{eqnarray}\mathbb{E}(\varphi_i(Y))\leq0,\hskip10pt \varphi_i:\,{\cal
Y}\rightarrow\mathbb{R},\;\;i=1,\ldots,d_{\varphi}.
\label{equation: moment inequalities}\end{eqnarray} This is a special case of our general
structure, where ${\cal U}\subseteq\mathbb{R}^{d_u}$ and
\[G(u)=\{y\in\mathcal{Y}:\;u_i\geq\varphi_i(y),\;\;i=1,\ldots,d_{u}\},\] and
$m_i(u)=u$, $i=1,\ldots,d_{\varphi}$, with
$d_u=d_{\varphi}$.\end{example}

\begin{example}
Model defined by conditional moment inequalities.
\begin{eqnarray}\mathbb{E}(\varphi_i(Y)\vert X)\leq0,\hskip10pt \varphi_i:\,{\cal
Y}\rightarrow\mathbb{R},\;\;i=1,\ldots,d_{\varphi},
\label{equation: conditional moment inequalities}\end{eqnarray}
where $X$ is a sub-vector of $Y$. \cite{Bierens:90} shows
that this model can be equivalently rephrased as
\begin{eqnarray}\mathbb{E}(\varphi_i(Y)1\{t_1\leq X\leq
t_2\})\leq0,\hskip10pt \varphi_i:\,{\cal
Y}\rightarrow\mathbb{R},\;\;i=1,\ldots,d_{\varphi},
\label{equation: transformed conditional moment
inequalities}\end{eqnarray} for all pairs
$(t_1,t_2)\in\mathbb{R}^{2d_x}$ (the inequality is understood
element by element). Conditionally on the observed sample, this
can be reduced to a finite set of moment inequalities by limiting
the class of pairs $(t_1,t_2)$ to observed pairs $(X_i,X_j)$,
$X_i<X_j$. Hence this fits into the framework of
example~\ref{example: moment inequalities}. \label{example:
conditional moment inequalities}\end{example}

\begin{example} Unobserved random censoring
(also known as accelerated failure time) model. A continuous
variable $Z=\mu(X)+\epsilon$, where $\mu$ is known, is censored by a random
variable $C$. The only observable variables are $X$, $V=\min(Z,C)$
and $D=1\{Z<C\}$. The error term $\epsilon$ is supposed to have
zero conditional median $P(\epsilon<0\vert X)=0$.
\cite{KT:2006} show that this model can be equivalently
rephrased in terms of unconditional moment inequalities.
\begin{eqnarray*}&&\mathbb{E}\left[\left(1\{V\geq\mu(X)\}-\frac{1}{2}\right)
1\{t_1\leq X\leq t_2\}\right]\leq0\\
&&\mathbb{E}\left[\left(\frac{1}{2}-D\times
1\{V\leq\mu(X)\}\right) 1\{t_1\leq X\leq
t_2\}\right]\geq0\end{eqnarray*} for all pairs
$(t_1,t_2)\in\mathbb{R}^{2d_x}$ (the inequality is understood
element by element). Hence this fits into the framework of
example~\ref{example: conditional moment inequalities}.
\label{example: random censoring}\end{example}

\begin{continued}[Example \protect\ref{example: social interactions} continued]
In case of models with multiple equilibria such as example~\ref{example: social interactions},
where the idiosyncratic disutility of smoking is only restricted by a finite collection of moment equalities, for instance $\mathbb{E}U=0$,
the model cannot be written in the familiar moment inequality formulation of example~\ref{example: moment inequalities}.
Hence the need for a new approach to the problem of falsifying the model.
\end{continued}

Finally we turn to an example of binary response, which we shall use as pilot
examples for illustrative purposes.

\begin{pilot} \label{pilot: binary response} {\bf A Binary Response
Model:} The observed variables $Y$ and $X$ are related by
$Z=1\{X+\varepsilon\leq0\}$, under the conditional median
restriction Pr$(\varepsilon\leq0\vert X)=\eta$ for a known $\eta$.
In our framework the vector of observed variables is $Y=(Z,X)'$,
and to deal with the conditioning, we take the vector $U$ to also
include $X$, i.e. $U=(X,\varepsilon)'$. To simplify exposition,
suppose $X$ only takes values in $\{-1,1\}$, so that ${\cal
Y}=\{0,1\}\times\{-1,1\}$ and $\;{\cal U}=\{-1,1\}\times[-2,2]$,
where the restriction on the domain of $\varepsilon$ is to ensure
compactness only. The multi-valued correspondence defining the
model is $\Gamma:{\cal U}\rightrightarrows{\cal Y}$
characterized by $\Gamma(1,x)=\{x\}\times(-2,-x]$
and $\Gamma(0,x)=\{x\}\times(-x,2]$. The two moment
restrictions are
$m_{\pm}(x,\varepsilon)=(1\{\varepsilon\leq0\}-\eta)(1\pm x)$.
\end{pilot}

We are now in the case where the economic model consists in the correspondence $G:\mathcal{U}\rightrightarrows
\mathcal{Y}$ and a finite set of moment restrictions on the distribution $\nu$ of unobservables.
Denote the model $(G,\mathcal{V})$.
Again, the observables are fully characterized
by their distribution $P$, which is unknown, but can be estimated from data.
Consider now the restrictions
imposed by the model on the joint distribution $\pi$ of the pair $(Y,U)$:
\begin{itemize} \item Its marginal with respect to $Y$ is $P$,
\item Its marginal with respect to $U$ belongs to $\mathcal{V}$, \item The
economic restrictions $Y\in G(U)$ hold $\pi$ almost
surely.
\end{itemize}
Again, a probability distribution $\pi$ that satisfies the restrictions above may or
may not exist. If and only if it does, we say that the distribution $P$ of
observable variables is compatible with the economic model $(G,\mathcal{V})$.

\begin{definition}\label{definition: semiparametric compatibility}
A distribution $P$ is compatible with the model $(G,\mathcal{V})$ for $(Y,U)$ if there exists a
law $\pi$ for the vector $(Y,U)$ with marginals $P$ with respect to $Y$ and marginal $\nu\in\mathcal{V}$
with respect to $U$ such that $\pi(\{Y\in G(U)\})=1$. \end{definition}

\subsection{Optimization formulation}
\label{subsection: optimization}

This hypothesis of compatibility
has a similar optimization interpretation as in the case of parametric restrictions on
unobservables. The distribution $P$ is compatible
with the model $(G,\mathcal{V})$ if and only if
$$\exists\pi\in{\cal M}(P,\mathcal{V}): \int_{\mathcal{Y}\times\mathcal{U}}1_{\{y\notin G(u)\}}d\pi(y,u)=0,$$
or equivalently \begin{eqnarray}\min_{\pi\in{\cal M}(P,\mathcal{V})} \int_{\mathcal{Y}\times\mathcal{U}}1_{\{y\notin G(u)\}}d\pi(y,u)=0.\label{equation: semiparametric min}\end{eqnarray}

Although this optimization problem differs from the optimal transportation problem
considered above, we shall see that inspection of the dual nevertheless provides a dimension reduction
which will allow to devise strategies to falsify the model based on a sample of realizations of $Y$.
However, before inspecting the dual, we need to show that the minimum in (\ref{equation: semiparametric min}) is actually attained, so that compatibility of observable distribution $P$ with the model $(G,\mathcal{V})$ is equivalent to
\begin{eqnarray}\inf_{\pi\in{\cal M}(P,\mathcal{V})} \int_{\mathcal{Y}\times\mathcal{U}}1_{\{y\notin G(u)\}}d\pi(y,u)=0.\label{equation: semiparametric inf}\end{eqnarray}

The following example shows that the
infimum is not always attained.

\begin{example}Let $P=N\left(0,1\right)$, $\mathcal{U}=\mathcal{Y}=
\mathbb{R}$, $\mathcal{V}=\left\{ \nu :
\mathbb{E}_{\nu}(U)=0\right\} $, and $\Gamma \left(
y\right) =\left\{ 1\right\} $ for all $y\in \mathcal{Y}$, and
consider the distribution $\pi _{m}=P\otimes \nu _{m}$ such that
$\nu _{m}\left( \left\{ 1\right\} \right) =1-1/m$, and $\nu
_{m}\left( \left\{ 1-m\right\} \right) = 1/m$. The $\pi_m$
probability of $Y\notin\Gamma(U)$ is $1/m$ which indeed
tends to zero as $m\rightarrow\infty$, but it is clear that there
exists no distribution $\nu$ which puts all mass on $\{1\}$ and
has expectation $0$.\label{example: quasi-consistent}\end{example}

It is clear from example~\ref{example: quasi-consistent} that we
need to make some form of assumption to avoid letting masses drift
off to infinity. The theorem below gives formal conditions under
which quasi-consistent alternatives are ruled out. It says
essentially that the moment functions $m(u)$ need to be
bounded.

\begin{assumption}[Uniform Integrability] \[\lim_{M\rightarrow \infty }\sup_{\nu\in\mathcal{V}}\nu \left[
\parallel\! m\left( U\right) \!\parallel 1_{\left\{ \parallel\! m\left(
U\right) \!\parallel >M\right\} }\right]
=0,\]
where $\parallel\! m\left( U\right) \!\parallel$
denotes the norm of the vector with components $m_i(U)$, for $1\leq i\leq d_m$.
\label{assumption: uniform integrability}\end{assumption}

\begin{assumption}[Tightness] For every $K\geq 0$, the set $\left\{ u:\parallel\!m\left(
u\right)\!\parallel \leq K\right\} $ is included in a
compact set.\label{assumption: tightness}\end{assumption}

Assumption~\ref{assumption: uniform integrability}
is an assumption of uniform integrability. It is immediate to note
that assumptions~\ref{assumption: uniform integrability}
and~\ref{assumption: tightness} are satisfied when the moment
functions $m(u)$ are bounded and $\mathcal{U}$ is
compact.

\begin{assumption}[Closed Graph] The graph of $G$, i.e. $\{(y,u)\in\mathcal{Y}\times\mathcal{U}:
y\in G(u)\}$ is closed. \label{assumption: closed
graph}\end{assumption}

In example~\ref{example: moment inequalities},
by Theorem~1.6 page 9 of \cite{RW:98}, we know that
assumption~\ref{assumption: closed graph} is satisfied when the
moment functions $\varphi_j$, $j=1,\ldots,d_{\varphi}$ are lower
semi-continuous.

We can now state the result:

\begin{theorem}
\label{theorem: quasi-consistent alternatives} Under
assumptions~\ref{assumption: uniform integrability},
\ref{assumption: tightness} and~\ref{assumption: closed graph},
(\ref{equation: semiparametric inf}) is equivalent to the compatibility of observable distribution $P$ with model $(G,\mathcal{V})$.
\end{theorem}

The two dual formulations of this optimization problem are the following:
\begin{eqnarray*}
&(\mbox{P})& \inf_{\pi\in{\cal M}(P,\mathcal{V})} \int_{{\cal Y}\times{\cal U}} 1_{\{y\notin G(u)\}}d\pi(y,u)\\
&(\mbox{D})& \sup_{f(y)+\lambda'm(u)\leq 1_{\{y\notin G(u)\}}} \int_{\cal Y} f\;dP.\end{eqnarray*}

Since $u$ does not enter in the dual functional, the dual constraint can be rewritten
as $f(y)=\inf_{u}\{1_{\{y\notin G(u)\}}-\lambda'm(u)\}$, so that the dual program can be rewritten
\[T(P,\mathcal{V}):=\sup_{\lambda\in\mathbb{R}^{d_m}}\int_\mathcal{Y} \left(\inf_{u\in\mathcal{U}}[1_{\{y\notin G(u)\}}-\lambda'm(u)]\right)dP(y),\]
which does not involve optimizing over an infinite dimensional space as the primal program did.

\begin{continued}[Pilot example \protect\ref{pilot: binary
response} continued] Here, we have
$\lambda=(\lambda_1,\lambda_2)\in \mathbb{R}^2$ and
\begin{eqnarray*}
g_{\lambda}(x,0)=\min (\inf_{\varepsilon\geq -x}\{
-\lambda'm(\varepsilon,x) \};\inf_{\varepsilon\leq -x}\{
1-\lambda'm(\varepsilon,x) \}),
\\g_{\lambda}(x,1)=\min (\inf_{\varepsilon\leq -x}\{
-\lambda'm(\varepsilon,x) \};\inf_{\varepsilon\geq -x}\{
1-\lambda'm(\varepsilon,x) \}).
\end{eqnarray*}
\end{continued}

However, the dual formulation
is useless if primal and dual are not equal. Note first that taking expectation in the dual constraint
immediately yields (D)$ \leq $(P), which is the weak duality inequality. The converse inequality is shown below.

\begin{assumption}[Slater Condition] There exists a $P$-integrable function $f$ and a vector $\lambda$ and
$\epsilon>0$ such that for all $(y,u)\in{\cal
Y}\times{\cal U}$, $f(y) + \lambda'm(u) <
1\{y\notin G(u)\}-\epsilon$.
\label{assumption: Slater}\end{assumption}

The Slater condition is an interior condition, i.e. it ensures there
exists a feasible solution to the optimization problem in the
interior of the constraints. Notice that when the $m_i$ are bounded, the Slater
condition is always satisfied.

\begin{theorem}[No Duality Gap]\label{theorem: inconsistent alternatives}
Under assumptions~\ref{assumption: uniform integrability},
\ref{assumption: tightness}, \ref{assumption: closed graph} and
\ref{assumption: Slater}, the observable distribution is compatible with model
$(G,\mathcal{V})$ if and only if $T(P,\mathcal{V})=0$.\label{theorem: semiparametric}\end{theorem}

As described in the appendix, this result is ensured by the fact
that there is no duality gap, i.e. that the statistic obtained by
duality is indeed positive when the primal is.

\subsection{Test of compatibility}
\label{subsection: semiparametric compatibility}

We now consider falsifiability of the model with semiparametric constraints on unobservables
through a test of the null hypothesis that $P$ is compatible with
$(G,\mathcal{V})$. Falsifying the model in this framework corresponds to the
finding that a sample $(Y_1,\ldots,Y_n)$ of $n$ copies
of $Y$ distributed according to the unknown true distribution $P$
was not been generated as part of an sample $((Y_1,U_1),\ldots,(Y_n,U_n))$
distributed according to a fixed $\pi$ with $U$-marginal $\nu$
in $\mathcal{V}$ and satisfying the restrictions $Y\in G(U)$ almost surely. Using the results of the previous section,
this can be expressed in the following equivalent ways.

\begin{proposition}
The following statements are equivalent:
\begin{eqnarray*}&\mbox{(i)}& \mbox{The observable distribution } P \mbox{ is compatible with the model } (G,\mathcal{V}),\\
&\mbox{(ii)}&\inf_{\pi\in{\cal M}(P,\mathcal{V})} \int_{{\cal Y}\times{\cal U}} 1_{\{y\notin G(u)\}}d\pi(y,u)=0,\\
&\mbox{(iii)}&\sup_{\lambda\in\mathbb{R}^{d_m}}\int_\mathcal{Y} \left(\inf_{u\in\mathcal{U}}[1_{\{y\notin G(u)\}}-\lambda'm(u)]\right)dP(y).\end{eqnarray*}\end{proposition}

Call $P_n$ the empirical distribution, defined by $P_n(A) = \sum_{i=1}^{n} 1_{Y_i\in A}/n$
for all $A$ measurable, and form the empirical analogues of the conditions above as
\begin{eqnarray*}
&(\mbox{EP})& \inf_{\pi\in{\cal M}(P_n,\mathcal{V})} \int_{{\cal Y}\times{\cal U}} 1_{\{y\notin G(u)\}}d\pi(y,u)\\
&(\mbox{ED})& \sup_{\lambda\in\mathbb{R}^{d_m}}\frac{1}{n}\sum_{i=1}^{n}\left(\inf_{u\in\mathcal{U}}[1_{\{Y_i\notin G(u)\}}-\lambda'm(u)]\right).\end{eqnarray*}
Note first that by the duality result of theorem~\ref{theorem: semiparametric}, the empirical primal (EP)
and the empirical dual (ED) are equal. As in the parametric case, the cost function $c(y,u)=1_{\{y\notin G(u)\}}$
can be replaced by $c(y,u)=d(y,G(u))>0$ if $y\notin G(u)$ and equal to $0$ if $y\in G(u)$, to yield a family of numerically
equivalent test statistics. Quantiles of their limiting distribution, or obtained from a bootstrap procedure can be used
to form a test of compatibility, however, since (ED) involves two consecutive optimizations, a computationally more appealing
procedure called dilation is proposed in \cite{GH:2006c}. The idea is to control the size of the test nonparametrically so as to compute (ED) only once. For a test with level $1-\alpha$, compute a correspondence $J_n:\mathcal{Y}\rightrightarrows\mathcal{Y}$ such that there exist a pair of random vectors $Y$ and $Y^\ast$ with
marginal distributions $P$ and $P_n$ respectively and satisfying $Y^\ast\in J_n(Y)$ with probability $1-\alpha$. The test then consists in rejecting compatibility of the unknown distribution $P$ of the observables with the model $(G,\mathcal{V})$ if and only if the known empirical distribution $P_n$ is not compatible with the model $(J_n\circ G,\mathcal{V})$, i.e. if
\[\sup_{\lambda\in\mathbb{R}^{d_m}}\frac{1}{n}\sum_{i=1}^{n}\left(\inf_{u\in\mathcal{U}}[1_{\{Y_i\notin J_n\circ G(u)\}}-\lambda'm(u)]\right)\ne0.\]

\section*{Conclusion}
We have proposed an optimal transportation formulation of the problem of testing compatibility
of an incompletely specified economic model with the distribution of its observable components.
In addition to relating this problem to a rich optimization literature, it allows the construction
of computable test statistics and the application of efficient combinatorial optimization algorithms
to the problem of inference in discrete games with multiple equilibria. A major application of tests of incomplete specifications is the construction of confidence regions for partially identified parameters. In this respect, the optimal transportation formulation proposed here allows the direct application of the methodology proposed in the seminal paper of \cite{CHT:2007} to general models with multiple equilibria.

\appendix

\section{Proof of results in the main text}

\begin{lemma}
\label{Lemma1} Under assumptions~\ref{assumption: uniform
integrability} and~\ref{assumption: tightness},
$\mathcal{V}$ is uniformly tight.
\end{lemma}

\begin{proof}[Proof of Lemma \protect\ref{Lemma1}]
For $M>1$, by assumptions~\ref{assumption: uniform integrability},%
\begin{eqnarray*}
\sup_{\nu\in{\cal V}}\nu \left( \left\{ \parallel\!
m\left( U\right) \!\parallel >M\right\} \right) \leq
\sup_{\nu\in{\cal V}}\nu \left[ \parallel\!
m(U)\!\parallel 1_{\left\{ \parallel\!
m(U)\!\parallel
>M\right\} }\right] \rightarrow 0\mbox{ as }M\rightarrow \infty ,
\end{eqnarray*}%
hence for $\epsilon >0$, there exists $M>0$ such that%
\begin{eqnarray*}
1-\epsilon \leq \sup_{\nu\in{\cal V}}\nu \left( \left\{
\parallel\! m\left( U\right) \!\parallel \leq M\right\} \right)
\end{eqnarray*}%
but by assumption~\ref{assumption: tightness}, there exists a
compact set $K$ such that $\left\{
\parallel\! m\left( U\right) \!\parallel \leq M\right\} \subset K$.\hskip4pt$\square$
\end{proof}

\begin{lemma}
\label{Lemma2}If $\mathcal{V}$ is uniformly tight, then $\mathcal{M}\left( P,%
\mathcal{V}\right) $ is uniformly tight.
\end{lemma}

\begin{proof}[Proof of Lemma \protect\ref{Lemma2}]
For $\epsilon >0$, there exists a compact $K_{Y}\subset
\mathcal{Y}$ such that $P\left( K_{Y}\right) \geq 1-\epsilon /2$;
by tightness of $\mathcal{V}$, there exists also a
compact $K_{U}\subset \mathcal{U}$ such that $\nu \left(
K_{U}\right) \geq 1-\epsilon /2$ for all $\nu \in
\mathcal{V}$. For every $\pi \in \mathcal{M}\left(
P,\mathcal{V}\right) $, one has $\pi \left( K_{Y}\times
K_{U}\right) \geq \max \left( P\left( K_{Y}\right) +\nu \left(
K_{U}\right) -1,0\right) $ (Fr\'{e}chet-Hoeffding lower bound),
thus $\pi \left( K_{Y}\times K_{U}\right) \geq 1-\epsilon
$.\hskip4pt$\square$
\end{proof}

\begin{proof}[Proof of Theorem \protect\ref{theorem: quasi-consistent alternatives}]
Suppose $\inf_{\pi \in {\cal M}(P,{\cal V})}E_{\pi
}\left[ 1\left\{ Y\notin G \left( U\right) \right\}
\right] =0$, we shall show that the infimum is actually attained.
Let $\pi _{n}\in {\cal M}\left( P,\mathcal{V}\right) $ a
sequence of
probability distributions of the joint couple $\left( U,Y\right) $ such that%
\begin{eqnarray*}
E_{\pi _{n}}\left[ 1\left\{ Y\notin G \left(
U\right) \right\} \right] \rightarrow 0.
\end{eqnarray*}

By Lemma \ref{Lemma2}, $\mathcal{M}\left(
P,\mathcal{V}\right) $ is uniformly tight, hence by
Prohorov's theorem it is relatively compact. Consequently there
exists a subsequence $\pi _{\varphi \left( n\right) }\in M\left(
P,\mathcal{V}\right) $ which is weakly convergent to $\pi
$.

One has $\pi \in \mathcal{M}\left( P,\mathcal{V}\right)$. Indeed, clearly $%
\pi _{Y}=P$, and by assumption~\ref{assumption: tightness} the sequences of random variables $%
m\left( U_{\varphi \left( n\right) }\right) $ are uniformly
integrable, therefore by \cite{Vaart:98}, Theorem 2.20, one
has $\pi _{\varphi \left( n\right) }\left[ m\left( U_{\varphi
\left( n\right) }\right) \right] \rightarrow \pi \left[
m\left( U\right) \right] $, thus $\pi \left[ m\left(
U\right) \right] =0$. Therefore, $\pi \in \mathcal{M}\left(
P,\mathcal{V} \right) $.

By assumption~\ref{assumption: closed graph}, the set $\left\{
Y\notin G \left( U\right) \right\} $ is open, hence
by the Portmanteau lemma (\cite{Vaart:98}, Lemma 2.2
formulation (v)),
\begin{eqnarray*}
\liminf \pi _{\varphi \left( n\right) }\left[ \left\{ Y\notin
G \left( U\right) \right\} \right] \geq \pi \left[
\left\{ Y\notin G\left( U\right) \right\} \right]
\end{eqnarray*}
thus $\pi \left[ \left\{ Y\notin G \left( U\right)
\right\} \right] =0$. \hskip4pt$\square$ \end{proof}

\begin{proof}[Proof of Theorem \protect\ref{theorem: inconsistent
alternatives}]

We need to show that the following two optimization problems $({\cal
P})$ and $({\cal P}^*)$ have finite solutions, and that they are
equal. \begin{eqnarray*} ({\cal P}): \hskip20pt
\sup_{(f,\lambda)\in{\cal C}^0\times \mathbb{R}^{d_m}} <P,f> \mbox{
subject to } Lf\leq\delta-\lambda'm
\end{eqnarray*} and \begin{eqnarray*} ({\cal P}^*): \hskip20pt \sup_{(\pi,\gamma)\in{\cal
M}\times \mathbb{R}^{d_m}} <\pi,\delta> \mbox{ subject to }
L^*\pi=P,\;\pi\geq0,\;<\pi,m>=0.
\end{eqnarray*}
where ${\cal C}^0$ is the space of continuous functions of $y$ and
$u$, equipped with the uniform topology, its dual with respect to
the scalar product $<Q,f>=\int f dQ$ is the space $\cal M$ of signed
(Radon) measures on ${\cal Y}\times{\cal U}$ equipped with the vague
topology (the weak topology with respect to this dual pair), $L$ is
the operator defined by $L(f)(y,u)=f(y)$ for all $u$, and its dual
$L^*$ is the projection of a measure $\pi$ on ${\cal Y}$, and the
function $\delta$ is defined by $\delta(y,u)=1\{y\notin G(u)\}$.
Note that $\delta(y,u)$ is not continuous, and hence is not included in
the dual of ${\cal M}$. However, since $G$ has a closed graph,
$\delta$ is lower semi-continuous, hence, so is the restriction
of the function \[<\delta,\pi>:=\int f d\pi\] to non-negative measures, and
the set of continuous functions such that $Lf\leq\delta$ is closed.

We now see that $({\cal P}^*)$ is the dual program of $({\cal P})$:
indeed, we have \begin{eqnarray*} &&\sup_{(f,\lambda)\in{\cal
C}^0\times \mathbb{R}^{d_m}} <P,f> \mbox{ subject to }
Lf\leq\delta-\lambda'm\\&=& \sup_{(f,\lambda)\in{\cal C}^0\times
\mathbb{R}^{d_m}} \inf_{\pi\geq0,\;\pi\in{\cal M}}
<P,f>+<\pi,\delta-\lambda'm-Lf>\\&=& \sup_{(f,\lambda)\in{\cal
C}^0\times \mathbb{R}^{d_m}} \inf_{\pi\geq0,\;\pi\in{\cal M}}
<P,f>+<\pi,\delta>-\lambda'<\pi,m>-<\pi,Lf>\\&=&
\sup_{(f,\lambda)\in{\cal C}^0\times \mathbb{R}^{d_m}}
\inf_{\pi\geq0,\;\pi\in{\cal M}}
<P,f>+<\pi,\delta>-\lambda'<\pi,m>-<L^*\pi,f>\\&=&
\sup_{(f,\lambda)\in{\cal C}^0\times \mathbb{R}^{d_m}}
\inf_{\pi\geq0,\;\pi\in{\cal M}}
<\pi,\delta>-\lambda'<\pi,m>+<P-L^*\pi,f>,
\end{eqnarray*} and \begin{eqnarray*}
&&\inf_{\pi\geq0,\;\pi\in{\cal M}}\sup_{(f,\lambda)\in{\cal
C}^0\times \mathbb{R}^{d_m}}
<\pi,\delta>-\lambda'<\pi,m>+<P-L^*\pi,f>\\&=&
\inf_{(\pi,\gamma)\in{\cal M}\times\mathbb{R}^{d_m}}<\pi,\delta>
\mbox{ subject to } <\pi,m>=0,\; L^*\pi=P,\;\pi\geq0.\end{eqnarray*}

We now proceed to prove that the strong duality holds, i.e. that the
infimum and supremum can be switched. Under condition
(\ref{assumption: Slater}), by Proposition (2.3) page 52 of
\cite{ET:76}, $({\cal P})$ is stable. Hence, by Proposition
(2.2) page 51 of \cite{ET:76}, $({\cal P})$ is normal and
$({\cal P}^*)$ has at least one solution. Finally, since
$f\mapsto<P,f>$ is linear, hence convex and lower semi-continuous,
by Proposition (2.1) page 51 of \cite{ET:76}, the two programs
are equal and have a finite solution.
\hskip4pt$\square$\end{proof}

\begin{acknowledgements}The three authors are also grateful to Victor Chernozhukov
and Pierre-Andr\'e Chiappori for many
helpful discussions (with the usual disclaimer).\end{acknowledgements}

\printbibliography

\end{document}